\documentclass[aps,prd,aps,floatfix,noshowpacs,tightenlines,noshowkeys,superscriptaddress,amsmath,amssymb,nofootinbib]{revtex4}
\usepackage{amssymb,amsbsy,epsfig,color,graphicx}
\usepackage{color}
\usepackage{[longtable}
\usepackage{array}
\usepackage{dcolumn}   
\usepackage{cellspace}
\usepackage{mathtools}
\usepackage{amstext}
\usepackage{amssymb}
\usepackage{stmaryrd}
\usepackage{stackrel}
\usepackage{graphicx}
\usepackage{esint}
\usepackage[utf8]{inputenc}
\usepackage{blindtext}
\usepackage{float}
\restylefloat{table}
\usepackage{booktabs}
\usepackage{enumitem} 

\usepackage{etoolbox} 
\usepackage{lipsum} 
\usepackage[capitalize]{cleveref}

\usepackage{multirow}
\usepackage[caption=false]{subfig}
\renewcommand\[{\begin{equation}}
\renewcommand\]{\end{equation}}
\newcommand{\al}{\alpha}
\newcommand{\bt}{\beta}

\newcommand{\ba}{\begin{eqnarray}}
\newcommand{\ea}{\end{eqnarray}}

\makeatletter

\appto{\appendix}{%
	\@ifstar{\def\theequation@prefix{A.}}%
	{}%
}
\makeatother

\begin{document}

\title{Towards nonsingular rotating compact object in ghost-free infinite derivative gravity}

	\author{Luca Buoninfante}
\affiliation{Dipartimento di Fisica "E.R. Caianiello", Universit\`a di Salerno, I-84084 Fisciano (SA), Italy}
\affiliation{INFN - Sezione di Napoli, Gruppo collegato di Salerno, I-84084 Fisciano (SA), Italy}
\affiliation{Van Swinderen Institute, University of Groningen, 9747 AG, Groningen, The Netherlands}

\author{Alan S. Cornell}
\affiliation{National Institute for Theoretical Physics; School of Physics, University of the Witwatersrand, Johannesburg, Wits 2050, South Africa}

\author{Gerhard Harmsen}
\affiliation{Van Swinderen Institute, University of Groningen, 9747 AG, Groningen, The Netherlands} 
\affiliation{National Institute for Theoretical Physics; School of Physics, University of the Witwatersrand, Johannesburg, Wits 2050, South Africa}

\author{Alexey S. Koshelev}
\affiliation{Departamento de F\'isica and Centro de Matem\'atica e Aplica\c c\~oes, Universidade da Beira Interior, 6200 Covilh\~a, Portugal}
\affiliation{Theoretische Natuurkunde, Vrije Universiteit Brussel and The International Solvay Institutes, Pleinlaan 2, B-1050, Brussels, Belgium}

\author{Gaetano Lambiase}
\affiliation{Dipartimento di Fisica "E.R. Caianiello", Universit\`a di Salerno, I-84084 Fisciano (SA), Italy}
\affiliation{INFN - Sezione di Napoli, Gruppo collegato di Salerno, I-84084 Fisciano (SA), Italy}

\author{Jo\~ao Marto}
\affiliation{Departamento de F\'isica and Centro de Matem\'atica e Aplica\c c\~oes, Universidade da Beira Interior, 6200 Covilh\~a, Portugal}

\author{Anupam Mazumdar}
\affiliation{Van Swinderen Institute, University of Groningen, 9747 AG, Groningen, The Netherlands}


\begin{abstract}
The vacuum solution of Einstein's theory of general relativity provides a rotating metric with a ring singularity, which is covered by the inner and outer horizons, and an ergo region. In this paper, we will discuss how ghost-free, quadratic curvature, Infinite Derivative Gravity (IDG) may resolve the ring singularity. In IDG the non-locality of the gravitational interaction can smear out the delta-Dirac source distribution by making the metric potential finite everywhere including at $r=0$. We show that the same feature also holds for a rotating metric. We can resolve the ring singularity such that no horizons are formed in the linear regime by smearing out a delta-source distribution on a ring. We will also show that the Kerr-metric does not solve the full non-linear equations of motion of ghost-free quadratic curvature IDG.
\end{abstract}

\maketitle

	
\section{Introduction}

Einstein's theory of general relativity (GR) is indeed a very successful metric theory of gravity which has seen amazing success in the infrared (IR)~\cite{-C.-M.}, including the 
detection of the first gravitational wave signal~\cite{-B.-P.}. Inspite of these successes, classical GR suffers from the ultraviolet (UV) catastrophe at short distances and small 
time scales, there are blackhole and cosmological singularities~\cite{Penrose:1964wq,Penrose:1969pc,Hawking:1969sw}. It has been recently shown that a ghost free, quadratic curvature infinite derivative gravity (IDG) can potentially resolve the cosmological~\cite{Biswas:2005qr}, and blackhole type singularities~\cite{Biswas:2011ar}. Infinite derivatives acting on a point Delta-Dirac source smears out the singularity by a Gaussian profile~\cite{Buoninfante:2018rlq,Tseytlin:1995uq,Siegel:2003vt}. At a quantum level the graviton vertex interactions become non-local~\cite{-Yu.-V.,Tomboulis,Tomboulis:2015gfa,Talaganis:2014ida}; very similar to string field theory~\cite{Witten:1985cc,Witten:2013pra,Woodard,Freund:1987kt}. Besides strings, non-locality is also a feature of loop quantum gravity, see~\cite{Rovelli:2011eq}, spin foam or dynamical triangulation where Wilson loops acts as fundamental operators, for a review~\cite{Ambjorn:2012jv}. The quantum scatterings for such non-local interactions in IDG provide a very interesting insight~\cite{Biswas:2014yia,Talaganis:2016ovm}, where there is a UV-IR connection when large number of scatterings of particles with non-local interactions are taken into account. The scattering amplitude gets exponentially suppressed for external momenta  $P^2>M_s^2$, and the scale of non-locality gets shifted by $M_s \rightarrow M_s/\sqrt{N}$~\cite{Buoninfante:2018mre}, for N-scatterers in the limit when $N\gg 1$. Furthermore, non-local thermal field theory provides resemblance to a Hagedorn phase as shown in~\cite{Biswas:2009nx}. 

Since, the gravitational interaction in the UV weakens,  both linear~\cite{Biswas:2011ar} and non-linear equations of motion~\cite{Biswas:2013cha} provide a conformally-flat spherically symmetric, static metric solution~\cite{Buoninfante:2018rlq}. A similar scenario also holds in the case of a charged point source \cite{Buoninfante:2018stt}. It has also been shown that the singularity and the event horizon does not form in a dynamical context at a linear level~\cite{Frolov:2015bia}, as a mass gap can be formed determined by the non-local scale \cite{Frolov}. In particular, it has been shown that singular solutions such as Schwarzschild metric~\cite{Koshelev:2018hpt}, Kasner metric~\cite{Koshelev:2018rau} do not satisfy the field equations in the vacuum. Moreover, infinite derivatives acting on the delta-Dirac distribution at the origin smears out by a Gaussian profile \cite{Biswas:2011ar,Buoninfante:2018rlq}, and the region of non-locality yields a non-vacuum solution as opposed to that in GR. 
 It is also possible to make the gravitational radius as large as the effective scale of non-locality, $r_{NL}$, which can be larger than the Schwarzschild's radius, $r_{sch} \leq r_{NL}$~\cite{Buoninfante:2018rlq}. At the cosmological front, such non-locality can potentially replace the cosmological singularity by big bounce~\cite{Biswas:2005qr} or freezing the Universe in the UV~\cite{Biswas:2006bs}. Outside the region of non-locality the gravitational interaction becomes that of GR, thus reproducing all the features of gravity being tested in the IR~\cite{Edholm:2016hbt}, similar features have been observed for extended objects such as d and p-branes~\cite{Boos:2018bxf}. In Ref. \cite{Giacchini:2018gxp}, it was shown that in higher curvature gravity with more than $4$ derivatives, the delta source gets smeared out, as for example in sixth order theory of gravity, and the linearized metric turns out to be singularity-free. However, such local theories always suffer from the presence of ghosts at the tree-level.

The aim of this paper will be to understand the rotating metric within IDG, and show how to  smear out the ring singularity of a Kerr metric~\cite{Kerr} in the linear regime by considering a toy-model with a delta-Dirac distribution on a rotating ring. We will show that the linear solution approaches conformal-flatness in the limit $r\rightarrow 0$. We will provide numerical/analytical solutions of the rotating metric and show how it recovers the predictions of GR in the IR. With the help of non-linear equations of motion we will show that  the Kerr metric does not pass through the field equations of ghost-free quadratic curvature IDG.


\section{The infinite covariant derivative action}

The most general quadratic curvature action, which is parity invariant and
torsion-free, is given by~\cite{Biswas:2011ar,Biswas:2013cha,Biswas:2016etb}
%
\begin{equation}
S=S_{EH}+S_q=\frac{1}{16\pi G}\int d^{4}x\sqrt{-g}\left[\mathcal{R}+\alpha_{c}\left(\mathcal{R}{\cal F}_{1}(\Box_{s})\mathcal{R}+\mathcal{R}^{\mu\nu}{\cal F}_{2}(\Box_{s})\mathcal{R}_{\mu\nu}+W^{\mu\nu\lambda\sigma}{\cal F}_{3}(\Box_{s})W_{\mu\nu\lambda\sigma}\right)\right]\,,\label{action}
\end{equation}
where  $S_{EH}$ corresponds to the Einstein-Hilbert and $S_q$ corresponds to the quadratic curvature terms, $G=1/M_{p}^{2}$ is Newton's gravitational constant, $\alpha_{c}\sim1/M_{s}^{2}$
is a dimensionful coupling, $\Box_{s}\equiv\Box/M_{s}^{2}$, where
$M_{s}$ represents the scale of non-locality at which new physics should emerge. In the limit $M_{s}\rightarrow\infty$,
the action reduces to the Einstein-Hilbert term, as expected. The d'Alembertian
operator is defined as $\Box=g^{\mu\nu}\nabla_{\mu}\nabla_{\nu}$, where $\mu\,,\nu=0,1,2,3$,
and we work with the mostly positive metric convention, $(-,+,+,+)$.
The three gravitational {\it form factors} ${\cal F}_{i}$'s, are analytic function of $\Box$ and can be expressed in series representation as follows
\begin{equation}\label{FF-exp}
{\cal F}_{i}(\Box_{s})=\sum_{n=0}^{\infty}f_{i,n}\Box_{s}^{n}\,,
\end{equation}
which are reminiscence to any massless theory possessing \textit{only} derivative
interactions. Note that we will always consider analytic operators of $\Box_s$, and not non-analytic operators such as $1/\Box_s$~\cite{Woodard-1,Conroy:2014eja}, or $\ln(\Box_s)$~\cite{Bravinsky}. The ghost-free condition around Minkowski background can be formulated as~\cite{Biswas:2011ar,Biswas:2013cha,Biswas:2016etb,Buoninfante:2016iuf}:
\begin{equation}\label{ghost-free condition}
6\mathcal{F}_1(\Box_s)+3\mathcal{F}_2(\Box_s)+2\mathcal{F}_3(\Box_s)=0\,,~~a(\Box_s)=1+2\mathcal{F}_2(\Box_s)\Box_s+4\mathcal{F}_3(\Box_s)\Box_s=e^{-\Box_s}\,.
\end{equation}
The field equations for the action in Eq.\eqref{action} have been derived in Ref. \cite{Biswas:2013cha}, and they are given by
\begin{align}
P^{\alpha\beta}= & -\frac{G^{\alpha\beta}}{8\pi G}+\frac{\alpha_{c}}{8\pi G}\biggl(4G^{\alpha\beta}{\cal F}_{1}(\Box_{s})\mathcal{R}+g^{\alpha\beta}\mathcal{R}{\cal F}_{1}(\Box_{s})\mathcal{R}-4\left(\nabla^{\alpha}\nabla^{\beta}-g^{\alpha\beta}\Box\right){\cal F}_{1}(\Box_{s})\mathcal{R}\nonumber \\
& -2\Omega_{1}^{\alpha\beta}+g^{\alpha\beta}(\Omega_{1\sigma}^{\;\sigma}+\bar{\Omega}_{1})+4\mathcal{R}_{\mu}^{\alpha}{\cal F}_{2}(\Box_{s})\mathcal{R}^{\mu\beta}\nonumber \\
& -g^{\alpha\beta}\mathcal{R}_{\nu}^{\mu}{\cal F}_{2}(\Box_{s})\mathcal{R}_{\mu}^{\nu}-4\nabla_{\mu}\nabla^{\beta}({\cal F}_{2}(\Box_{s})\mathcal{R}^{\mu\alpha})+2\Box({\cal F}_{2}(\Box_{s})\mathcal{R}^{\alpha\beta})
\nonumber \\
& +2g^{\alpha\beta}\nabla_{\mu}\nabla_{\nu}({\cal F}_{2}(\Box_{s})\mathcal{R}^{\mu\nu})-2\Omega_{2}^{\alpha\beta}+g^{\alpha\beta}(\Omega_{2\sigma}^{\;\sigma}+\bar{\Omega}_{2})-4\Delta_{2}^{\alpha\beta}\nonumber \\
& -g^{\alpha\beta}W^{\mu\nu\lambda\sigma}{\cal F}_{3}(\Box_{s})W_{\mu\nu\lambda\sigma}+4W_{\;\mu\nu\sigma}^{\alpha}{\cal {\cal F}}_{3}(\Box_{s})W^{\beta\mu\nu\sigma}\nonumber \\
& -4(\mathcal{R}_{\mu\nu}+2\nabla_{\mu}\nabla_{\nu})({\cal {\cal F}}_{3}(\Box_{s})W^{\beta\mu\nu\alpha})-2\Omega_{3}^{\alpha\beta}+g^{\alpha\beta}(\Omega_{3\gamma}^{\;\gamma}+\bar{\Omega}_{3})-8\Delta_{3}^{\alpha\beta}\biggr)\nonumber \\
= & -T^{\al\bt}\,,\label{EOM}
\end{align}
where $T^{\al\bt}$ is the stress-energy tensor of the matter component, and the symmetric tensors are defined as (see~Ref. \cite{Biswas:2013cha}): 
\begin{align}
\Omega_{1}^{\alpha\beta}= & \sum_{n=1}^{\infty}f_{1_{n}}\sum_{l=0}^{n-1}\nabla^{\alpha}\mathcal{R}^{(l)}\nabla^{\beta}\mathcal{R}^{(n-l-1)},\quad\bar{\Omega}_{1}=\sum_{n=1}^{\infty}f_{1_{n}}\sum_{l=0}^{n-1}\mathcal{R}^{(l)}\mathcal{R}^{(n-l)},\label{details}\\
\Omega_{2}^{\alpha\beta}= & \sum_{n=1}^{\infty}f_{2_{n}}\sum_{l=0}^{n-1}\mathcal{R}_{\nu}^{\mu;\alpha(l)}\mathcal{R}_{\mu}^{\nu;\beta(n-l-1)},\quad\bar{\Omega}_{2}=\sum_{n=1}^{\infty}f_{2_{n}}\sum_{l=0}^{n-1}\mathcal{R}_{\nu}^{\mu(l)}\mathcal{R}_{\mu}^{\nu(n-l)}\,,\\
\Delta_{2}^{\alpha\beta}= & \sum_{n=1}^{\infty}f_{2_{n}}\sum_{l=0}^{n-1}[\mathcal{R}_{\sigma}^{\nu(l)}\mathcal{R}^{(\beta\sigma;\alpha)(n-l-1)}-\mathcal{R}_{\;\sigma}^{\nu;\alpha(l)}\mathcal{R}^{\beta\sigma(n-l-1)}]_{;\nu}\,,\\
\Omega_{3}^{\alpha\beta}= & \sum_{n=1}^{\infty}f_{3_{n}}\sum_{l=0}^{n-1}W_{\:\:\nu\lambda\sigma}^{\mu;\alpha(l)}W_{\mu}^{\;\nu\lambda\sigma;\beta(n-l-1)},\quad\bar{\Omega}_{3}=\sum_{n=1}^{\infty}f_{3_{n}}\sum_{l=0}^{n-1}W_{\:\:\nu\lambda\sigma}^{\mu(l)}W_{\mu}^{\;\nu\lambda\sigma(n-l)}\,,\label{details-1}\\
\Delta_{3}^{\alpha\beta}= & \sum_{n=1}^{\infty}f_{3_{n}}\sum_{l=0}^{n-1}[W_{\quad\sigma\mu}^{\lambda\nu(l)}W_{\lambda}^{\;\beta\sigma\mu;\alpha(n-l-1)}-W_{\quad\sigma\mu}^{\lambda\nu\;\;;\alpha(l)}W_{\lambda}^{\:\beta\sigma\mu(n-l-1)}]_{;\nu}\,.\label{details-2}
\end{align}
The notation $\mathcal{R}^{(l)}\equiv\Box^{l}\mathcal{R}$ is only used for the covariant derivatives acting on the curvature tensors. We can also compute the trace of the field equations in Eq.\eqref{EOM} whose expression is a lot simpler and is given by~\cite{Biswas:2013cha}:
\begin{align}
P= & \frac{\mathcal{R}}{8\pi G}+\frac{\alpha_{c}}{8\pi G}\biggl(12\Box{\cal F}_{1}(\Box_{s})\mathcal{R}+2\Box({\cal F}_{2}(\Box_{s})\mathcal{R})+4\nabla_{\mu}\nabla_{\nu}({\cal F}_{2}(\Box_{s})\mathcal{R}^{\mu\nu})\nonumber \\
& +2(\Omega_{1\sigma}^{\;\sigma}+2\bar{\Omega}_{1})+2(\Omega_{2\sigma}^{\;\sigma}+2\bar{\Omega}_{2})+2(\Omega_{3\sigma}^{\;\sigma}+2\bar{\Omega}_{3})-4\Delta_{2\sigma}^{\;\sigma}-8\Delta_{3\sigma}^{\;\sigma}\biggr)\nonumber \\
= & -T\equiv -g_{\al\bt}T^{\al\bt}\,.\label{trace}
\end{align}
The static solution for both linerized~\cite{Biswas:2011ar,Buoninfante:2018xiw,Koshelev:2017bxd}, and the full non-linear regime~\cite{Koshelev:2018hpt,Buoninfante:2018rlq} have shown that Schwarzschild-like singular solution is not permissible within IDG. In the UV, well inside the region of non-locality, $r \ll 1/M_s$, the Weyl tensor  $W^{\mu\nu\lambda\sigma}\rightarrow 0$ as $r\rightarrow 0$.  In this respect, the system has some similarity with fuzz-ball~\cite{Mathur}. The smearing out of the singularity has also been seen in non-commutative geometry, as pointed out first in Ref.~\cite{Nicolini}. In GR the Schwarzschild metric is derived by imposing the boundary condition at the origin, i.e. by putting a delta-Dirac distribution at $r=0$ \cite{basalin1,basalin2}, in our case, the IDG smears this singularity at the origin.  The entire spacetime metric is regular in the static case, inside the non-local region, i.e. $r\ll 2/M_s$, without any singularity. Therefore, perturbation theory can be trusted all the way from $r=0$ to $r\rightarrow \infty$ as long as $mM_s<M_p^2;$ see Ref.\cite{Biswas:2011ar,Buoninfante:2018rlq}. Since, the effective scale of non-locality is given by  $M_s \rightarrow M_s/\sqrt{N}$, where $N$ is number of graviton interacting non-locally,  the condition, $mM_s<M_p^2$, can be satisfied for large astrophysical mass $m$~\cite{Koshelev:2017bxd}.


\section{Ring singularity}

Let us briefly recall the Kerr metric~\cite{Kerr} in rational polynomial coordinates, which is given by, see~\cite{Visser:2007fj}:
\begin{equation}
\begin{aligned}ds^{2}= & -\left(1-\dfrac{2mr}{r^{2}+a^{2}\chi^{2}}\right)dt^{2}-\dfrac{4mar\left(1-\chi^{2}\right)}{r^{2}+a^{2}\chi^{2}}dtd\varphi+\dfrac{r^{2}+a^{2}\chi^{2}}{r^{2}-2mr+a^{2}}dr^{2}+\left(r^{2}+a^{2}\chi^{2}\right)\dfrac{d\chi^{2}}{1-\chi^{2}}\\
& +\left(1-\chi^{2}\right)\left(r^{2}+a^{2}+\dfrac{2ma^{2}r\left(1-\chi^{2}\right)}{r^{2}+a^{2}\chi^{2}}\right)d\varphi^{2},
\end{aligned}
\label{metric-1}
\end{equation}
where $\chi=\cos\theta$ is the transformation used to bring the standard
Boyer-Lindquist coordinates, while $m$ is the mass and $J=am$ is the angular momentum, with $a$ being the rotation parameter. One of the key observation is that the Kerr metric has a ring-singularity which is described by the equation (see Ref. \cite{Hawking:1973uf} for a nice discussion):
%
$r^2+a^2{\rm cos}^2\theta=0$,
%
where it is clear that $a$ corresponds to the radius of the ring, while $r$ is the radial coordinate in Boyer-Lindsquit coordinates, which are defined in terms of the Cartesian ones as follows: $x=\sqrt{r^2+a^2}~{\rm sin}\theta{\rm cos}\varphi,$
$y=\sqrt{r^2+a^2}{\rm sin}\theta{\rm sin}\varphi $, and $z=r{\rm cos}\theta$.
The Kretschmann scalar blows up when $r^2+a^2{\rm cos}^2\theta=0$ is satisfied, i.e. when $r=0$ and $\theta=\pi/2$, or in Cartesian coordinates,
%
$x^2+y^2=a^2,\,z=0$,
%
namely the ring singularity lies on a plane which is perpendicular to the rotation axis. 
Let us first discuss the physics in the linear regime, in analogy with the static case~\footnote{There were attempts to understand the Kerr metric in IDG, see~\cite{Cornell:2017irh}. However, we have found an error in our analysis, which we have rectified here. Unfortunately, the rotation was not taken into account correctly in the paper.}. We consider the source is a delta-Dirac distribution on a ring of radius $a$, which is rotating with a constant angular velocity, $\omega$, in the plane $x$-$y$ $(z=0).$ Thus, the components of the energy momentum tensor of the source are:
\begin{equation}
T_{00}=m\delta(z)\frac{\delta(x^2+y^2-a^2)}{\pi}\,,~~~~ T_{0i}=T_{00} v_i. \label{ring}
\end{equation}
Note that the factor $\pi$ in the denominator comes from the fact that: $\delta(x,y)\equiv \delta(x)\delta(y)=\pi \delta(x^2+y^2)$,  and $v_i$ is the tangential velocity whose magnitude can be expressed as $v=\omega a$, and assuming that the rotation happens around the $z$-axis, we have $v_x=-y \omega,$ $v_y=x \omega,$ $v_z=0$. Note that this choice of the stres-energy tensor, in analogy with the static case, is compatible with the fact that in order for the Einstein equations and the Kerr metric to be defined in the entire spacetime we need a non-vanishing stress-energy tensor at the ring. In fact, by using the theory of distribution \cite{basalin2}, it was rigorously shown that the stress-energy tensor for a Kerr metric has a structure similar to the one we have written in Eq.\eqref{ring}. For example, the $(00)$-component of the Einstein tensor in the case of the Kerr metric is $G_{00}\sim m\delta(z)\delta(x^2+y^2-a^2)$ \cite{basalin2}. A general linearized metric, which can describe the spacetime in presence of a rotating source can be written, in isotropic coordinates, as
\begin{equation}
ds^2=-(1+2\Phi)dt^2+2 \vec{h}\cdot d\vec{x}dt+(1-2\Psi)d\vec{x}^2,\label{metric}
\end{equation}
where $h_{00}=-2\Phi < 1$, $h_{ij}=-2\Psi\delta_{ij} < 1$ and $h_{0i}\equiv h_i < 1$ signify the weak-field and the slow rotation regime, and now the metric components depend on the isotropic radius, $r$, which should not be confused with the Boyer-Lindsquit radial coordinate used above. To find the form of the metric components, we would need to solve the following differential equations:
\begin{equation}
\begin{array}{rl}
\displaystyle e^{-\nabla^2/M_s^2}\nabla^2 \Phi(\vec{r}) = \displaystyle e^{-\nabla^2/M_s^2}\nabla^2 \Psi(\vec{r}) = & 4Gm \delta(z)\delta(x^2+y^2-a^2),\\
\displaystyle e^{-\nabla^2/M_s^2}\nabla^2 h_{0x}(\vec{r})=& \displaystyle-16Gm\omega y \delta(z)\delta(x^2+y^2-a^2),\\
\displaystyle e^{-\nabla^2/M_s^2}\nabla^2 h_{0y}(\vec{r})=& \displaystyle 16Gm\omega x \delta(z)\delta(x^2+y^2-a^2);
\end{array}\label{diff-eq}
\end{equation}
where we are assuming the ghost-free condition in Eq.\eqref{ghost-free condition}.
To solve the differential equations in Eq.\eqref{diff-eq} we can go to the Fourier space and then anti-transform back to coordinate space; thus, first of all, we need to compute the Fourier transforms of the stress-energy tensor components, i.e. of $T_{00}$ and $T_{0i}.$


\subsection{Smearing out the ring singularity at the linearized level}

Let us first compute the corresponding gravitational potential, $\Phi=\Psi$;  we need the Fourier transform of the ring-distribution in Eq.\eqref{ring}:
\begin{equation}
\mathcal{F}[\delta(z)\delta(x^2+y^2-a^2)]=\displaystyle \int dxdydz \delta(z)\delta(x^2+y^2-a^2) e^{ik_xx}e^{ik_yy}e^{ik_zz}.
\end{equation}
It can be computed by performing the integral in cylindrical coordinates: $x=\rho {\cos}\varphi,$ $y=\rho {\rm sin}\varphi,$ $z=z$, so that
\begin{equation}
\mathcal{F}[\delta(z)\delta(x^2+y^2-a^2)]=  \displaystyle\int\limits_{-\infty}^{\infty} dz \delta(z) e^{ik_z z}\int\limits_{0}^{\infty}d\rho \rho \delta(\rho^2-a^2)  \int\limits_{0}^{2\pi}d\varphi  e^{ik_x\rho {\cos}\varphi}e^{ik_y\rho {\rm sin}\varphi}
= \displaystyle \pi I_0\left(ia\sqrt{k_x^2+k_y^2}\right),\label{fourier00}
\end{equation}
where $I_0$ is a Modified Bessel function, which is also defined in terms of the Bessel function as $I_0(x)=J_0(ix)$.
\begin{figure}[t!]
	\includegraphics[scale=0.45]{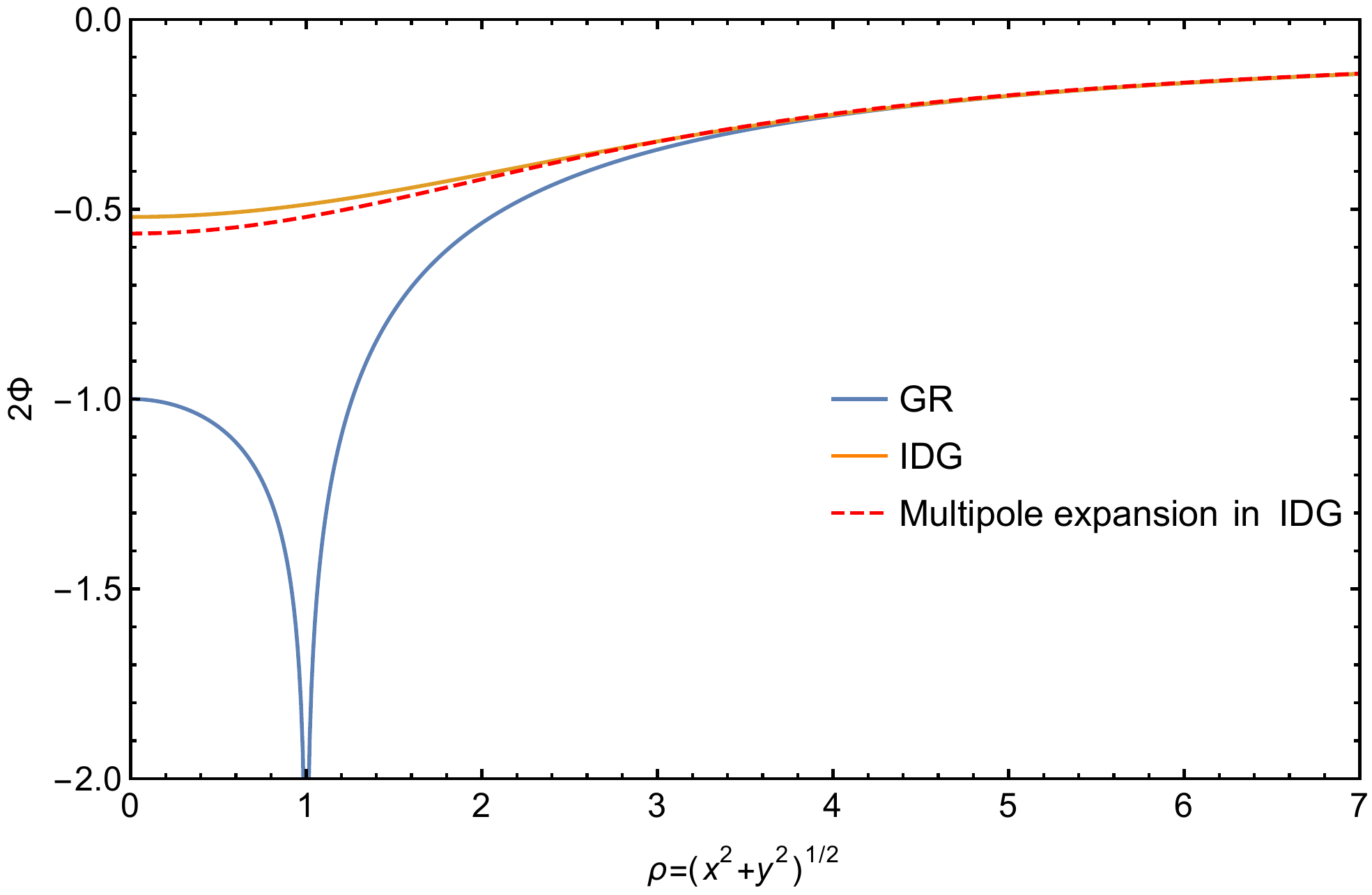}
	\centering
	\protect\caption{In this plot we have shown the results of the numerical computation for the integrals in Eq.\eqref{grav-potIDGnum}, and the behavior of the metric potential in the case of the multipole expansion in Eq.\eqref{metric-IDG}. The blue line corresponds to the behavior of the perturbation, $2\Phi=-h_{00}$, in GR; the orange line to the behavior of the metric potential in IDG; while the dashed red line represents the metric potential in the case of the multipole expansion. We have chosen $G=1,$ $m=0.5,$ $a=1$ and $M_s=0.9.$ We can notice that the gravitational potential in GR blows up for $\rho=a=1,$ while it is finite in IDG; moreover, the metric coming from the multipole expansion is a very good approximation outside the source, i.e. for $\rho>a.$ }
\end{figure}
By anti-transforming, we obtain the gravitational potential in coordinate space:
\begin{equation}
\Phi(\vec{r})= -4\pi Gm \int \frac{d^3k}{(2\pi)^3}\frac{e^{-k^2/M_s^2}}{k^2} I_0\left(ia\sqrt{k_x^2+k_y^2}\right)e^{ik_xx}e^{ik_yy}e^{ik_zz}, \label{grav-pot}
\end{equation}
where $d^3k=dk_xdk_ydk_z$ and $k^2=k_x^2+k_y^2+k_z^2.$ In order to study whether the ring singularity is still present or not in IDG, we can simplify the integral in Eq.\eqref{grav-pot}, by confining ourselves on the $x$-$y$ ($z=0$) plane, where the ring singularity lies in the case of Kerr metric.
Thus, by setting $z=0$, and going to cylindrical coordinates, $k_x=\zeta{\rm cos}\varphi,$ $k_y=\zeta{\rm sin}\varphi$, $k_z=k_z$, we can rewrite the integral in Eq.\eqref{grav-pot} as follows:
\begin{equation}
\Phi(\rho)= -Gm \int\limits_{0}^{\infty}d\zeta I_0\left(ia\zeta\right)I_0\left(i\zeta \rho \right){\rm Erfc}\left(\frac{\zeta}{M_s}\right),~~~~
{\rm when}~M_s\rightarrow \infty \Longrightarrow~\Phi_{GR}(\rho)= - Gm \int\limits_{0}^{\infty}d\zeta I_0\left(ia\zeta\right)I_0\left(i\zeta\rho\right),
\label{grav-potIDGnum}
\end{equation}
where the last integral corresponds to the GR case.
%
%
The two integrals in Eq. \eqref{grav-potIDGnum} cannot be solved analytically, but we can compute them numerically. From the numerical computation one can explicitly see that for, $x^2+y^2=a^2$, the gravitational potential in GR diverges as expected, while in IDG it is singularity-free; see Fig. 1. This is what we have expected to be physically; the IDG smears out a ring distribution very similarly to the case of a point source~\cite{Biswas:2005qr,Biswas:2011ar,Buoninfante:2018rlq,Buoninfante:2018xiw}.
Furthermore, we can trust the linear regime all the way up to $\rho=0$, as long as $2\Phi(0)<1.$ The integral in Eq.\eqref{grav-potIDGnum} can be evaluated analytically at $\rho=0$:
\begin{equation}
\Phi(0)=-\frac{Gm}{a}{\rm Erf}\left(\frac{M_sa}{2}\right)\,;
\end{equation}
the linearized approximation yields $2\Phi(0)<1$. Since ${\rm Erf}\left({M_sa}/{2}\right)<1$, the case $a>2Gm$ always satisfies the inequality; while in the opposite case $a<2Gm$, the weak-field inequality is satisfied as long as
\begin{equation}
a < \frac{2}{M_s}~~({\rm radius~of~the~ring}<{\rm scale~of~non-locality})\,. \label{a-ineq}
\end{equation}
This suggests that ghost-free IDG can indeed avoid the ring-type singularity. 


\subsection{Computing $h_{0i}$ components for a rotating ring }

So far we have only computed the static gravitational potential generated by a delta-Dirac distribution on the ring. We now wish to study the components $h_{0i}$ which are related to the fact that the ring is also rotating with a constant angular velocity $\omega$. We would need to compute the following Fourier transform:
\begin{equation}
\mathcal{F}[j~\delta(z)\delta(x^2+y^2-a^2)]=\displaystyle \int dxdydz~j~\delta(z)\delta(x^2+y^2-a^2) e^{ik_xx}e^{ik_yy}e^{ik_zz},
\end{equation}
%
%
where $j=x,~ y$. The computation can be performed by using cylindrical coordinates as done in Eq.\eqref{fourier00}:
\begin{equation}
\begin{array}{rl}
\mathcal{F}[x~\delta(z)\delta(x^2+y^2-a^2)]= & \displaystyle\int\limits_{-\infty}^{\infty} dz \delta(z) e^{ik_z z}\int\limits_{0}^{\infty}d\rho \rho^2  \delta(\rho^2-a^2)  \int\limits_{0}^{2\pi}d\varphi  e^{ik_x\rho {\cos}\varphi}e^{ik_y\rho {\rm sin}\varphi}{\rm cos}\varphi\\
=& \displaystyle \pi  a \frac{k_x}{\sqrt{k_x^2+k_y^2}} I_1\left(ia\sqrt{k_x^2+k_y^2}\right),
\end{array}
\end{equation}
and by following similar steps we also obtain:
\begin{equation}
\mathcal{F}[y\delta(z)\delta(x^2+y^2-a^2)]=\pi a \frac{k_y}{\sqrt{k_x^2+k_y^2}} I_1\left(ia\sqrt{k_x^2+k_y^2}\right),
\end{equation}
where $I_1$ is a Modified Bessel function. We can express the components $h_{0j}$ in coordinate space as anti-transforms:
\begin{equation}
h_{0j}(\vec{r})= 16Gm\omega a \int \frac{d^3k}{(2\pi)^3}\frac{e^{-k^2/M_s^2}}{k^2} \frac{k_j}{\sqrt{k_x^2+k_y^2}}I_1\left(ia\sqrt{k_x^2+k_y^2}\right)e^{ik_xx}e^{ik_yy}e^{ik_zz}, \label{int-h0x}
\end{equation}
%
%
where $j=x,~y$. By using cylindrical coordinates, similar to the integrands in Eq.\eqref{grav-potIDGnum}, and setting $z=0,$ we obtain similar expressions for the cross-terms:
\begin{equation}
h_{0x}(x,y)= 4Gm\omega a \frac{y}{\rho}\int\limits_{0}^{\infty} d\zeta I_1(ia\zeta)I_1(i\zeta \rho) {\rm Erfc}\left(\frac{\zeta}{M_s}\right)\,,~~
%
h_{0y}(x,y)= -4Gm\omega a \frac{x}{\rho}\int\limits_{0}^{\infty} d\zeta I_1(ia\zeta)I_1(i\zeta \rho) {\rm Erfc}\left(\frac{\zeta}{M_s}\right),\label{h0y}
\end{equation}
where $\rho=\sqrt{x^2+y^2}$ is the radial cylindrical coordinate in the plane $z=0.$
\begin{figure}[t!]
	\includegraphics[scale=0.45]{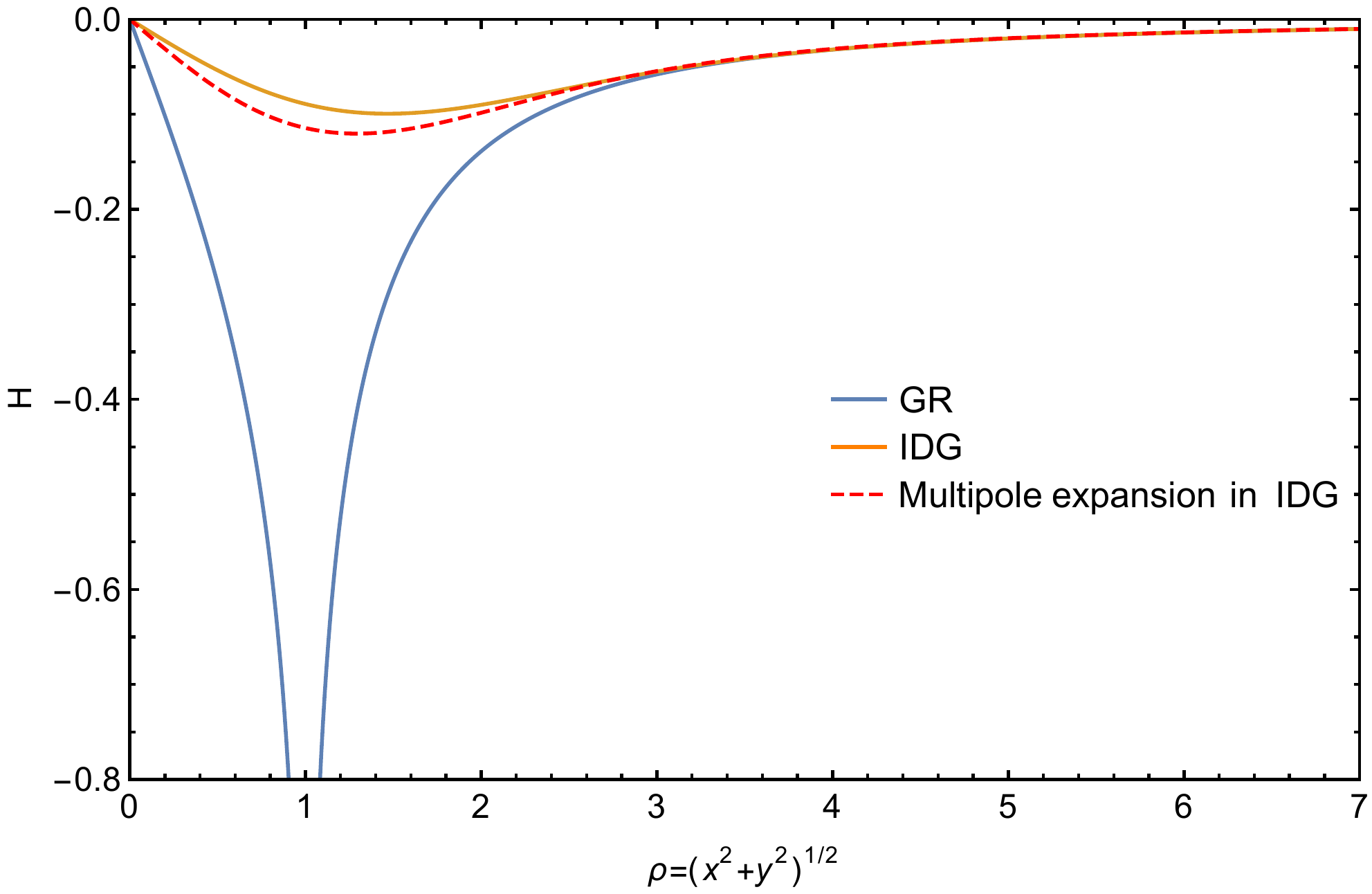}
	\centering
	\protect\caption{In this plot we have shown the results of the numerical computation for the integrals in Eq. \eqref{H-IDG} and the behavior of the same function in the case of the multipole expansion in Eq.\eqref{coord-trasf-IDG}. The blue line corresponds to the behavior of the function $H_{GR}$, and so of the cross-term in GR; the orange line to the behavior of the function $H_{IDG}$, and so of the cross-term in IDG; while the dashed red line represents the cross-term in the case of the multipole expansion. For convenience we have chosen $a=1$ and $M_s=1.5.$ We can notice that the metric components $h_{0i}$ blow up in GR for $\rho=a=1,$ while they are finite in IDG; moreover, the metric coming from the multipole expansion is a very good approximation outside the source, i.e. for $\rho>a.$}
\end{figure}
Since $\theta=\pi/2$, we have
%
$x=\rho{\rm cos}\varphi,\,\, y=\rho {\rm sin}\varphi $,
%
thus all the radial dependence and the singularity structure are taken into account by the integrals:
\begin{equation}
H(\rho):=\int\limits_{0}^{\infty} d\zeta I_1(ia\zeta)I_1(i\zeta \rho) {\rm Erfc}\left(\frac{\zeta}{M_s}\right),~~
{\rm when}~M_s\rightarrow \infty~\Longrightarrow~H_{GR}(\rho):=\int\limits_{0}^{\infty} d\zeta I_1(ia\zeta)I_1(i\zeta \rho)
\label{H-IDG}
\end{equation}
where the last integral corresponds to the GR case.
%
%
The two integrals in Eq.\eqref{H-IDG} cannot be solved analytically but we can compute them numerically and check the absence of any singularities. As it also happens for the potentials $h_{00}$ and $h_{ij}$, the cross-term $h_{0i}$ show the presence of a ring singularity in GR; indeed, from the numerical analysis one can explicitly see that for $x^2+y^2=a^2$ the function $H_{GR}$ diverges in GR. While in IDG the cross-term turns out to be singularity-free; indeed, the function $H$ is finite everywhere. 
In analogy with the static scenario, also in the rotating case, IDG is responsible for a smearing effect, in this case of the delta-Dirac ring distribution. Note that at the origin, $\rho=0,$ $z=0,$ the cross-term vanishes, which implies that in IDG the spacetime metric approaches conformal-flatness; indeed, at $r=0$ the rotating metric becomes that of the static case~\cite{Buoninfante:2018xiw}. 

In the IR regime, for $\rho\gg a$, the metric components found above match extremely well with the case of GR. Indeed, for distances larger than the radius of the ring and the scale of non-locality, i.e. $\rho\gg 2/M_s>a$, we recover the Lense-Thirring metric~\cite{Lense}\footnote{Recall that the Lense-Thirring metric represents the weak-field and slow-rotation limit of the Kerr metric \cite{Lense}:
	\begin{equation}
	ds^2=-\left(1-\frac{2Gm}{r}\right)dt^2+\frac{4GJ}{r^3}(ydxdt-xdydt)+\left(1+\frac{2Gm}{r}\right)(dr^2+r^2d\Omega^2).\label{lense-thirr}
	\end{equation}
}.
%
%
%
%
%
%
To exactly recover the Lense-Thirring metric at large distances, we need to identify $J=ma^2\omega$, which is nothing but the relation $J=I\omega,$ where $I=ma^2$ is the moment of inertia of the delta-Dirac ring distribution. Note that  the relation $J=am$ does not hold, but the angular momentum is related to the parameter $a$ through the momentum of inertia of the source.


\subsection{Rotating metric outside the source: multipole expansion in IDG}

We now wish to determine the generic form of the metric in IDG outside the rotating source, without assuming any large distance limit. In this regime, the linear treatment is valid, see Eq.\eqref{metric}. The components $h_{00}$ and $h_{ij}$ will be the same already obtained in the static case; while to compute the $(0i)$-components we can consider a multipole expansion for ${\rm Erf}\left(M_s|\vec{r}-\vec{r}'|/2\right)/|\vec{r}-\vec{r}'|$, 
\begin{equation}
\frac{1}{|\vec{r}-\vec{r}'|}{\rm Erf}\left(\frac{M_s|\vec{r}-\vec{r}'|}{2}\right)=\frac{1}{r}{\rm Erf}\left(\frac{M_sr}{2}\right)+\left[\frac{1}{r^3}{\rm Erf}\left(\frac{M_sr}{2}\right)-\frac{M_s}{\sqrt{\pi}r^2}e^{-\frac{M_s^2r^2}{4}}\right]\sum\limits_{j=1}^{3}x^jx'^j+\cdots;\label{multipole-IDG}
\end{equation}
which recovers the GR case in the large distance regime, $M_sr\gg2$, as expected. Such a multipole expansion holds true for $r>r'\sim a,$ which means outside the source.
By using Eq.\eqref{multipole-IDG},  we can now compute the $h_{0i}$ components 
\begin{equation}
h_{0i}(\vec{r})=  \displaystyle 4 G\int d^3r' \frac{T_{0i}(\vec{r}')}{|\vec{r}-\vec{r}'|}{\rm Erf}\left(\frac{M_s|\vec{r}-\vec{r}'|}{2}\right)
=  \displaystyle 2G\left[\frac{1}{r^3}{\rm Erf}\left(\frac{M_sr}{2}\right)-\frac{M_s}{\sqrt{\pi}r^2}e^{-\frac{M_s^2r^2}{4}}\right](\vec{r}\times \vec{J})_i,
\end{equation}
We can move from Cartesian to isotropic coordinates, so that the $d\varphi dt$ component of the metric will be given by:
\begin{equation}
\begin{array}{rl}
2 \vec{h}\cdot d\vec{x}dt
= & \displaystyle -4GJ\left[\frac{1}{r}{\rm Erf}\left(\frac{M_sr}{2}\right)-\frac{M_s}{\sqrt{\pi}}e^{-\frac{M_s^2r^2}{4}}\right]{\rm sin}^2\theta d\varphi dt,
\end{array}\label{coord-trasf-IDG}
\end{equation}
which in the regime $M_sr\gg2$ recovers GR result, as expected.  Moreover, by expressing $J=I\omega=ma^2\omega$ and imposing $|h_{0i}|\sim GmM_s^2\omega a^2<1,$ we can notice that the slow rotation regime means $\omega<1/a,$ when we also require $GmM_s<1$ and $aM_s<1.$ Note also that by recasting the cross-term in terms of the angular momentum and imposing the linearized regime we obtain $|h_{0i}|\sim GM_s^2J<1$, which also means; $J<(M_p/M_s)^2.$ From the last inequality, since $M_s\leq M_p,$  the angular momentum $J$ may also exceed one in IDG. The linearized spacetime metric in Eq.\eqref{metric} outside the source, $ r > a$, in the case of IDG reads:
\begin{equation}
\begin{array}{rl}
ds^2=& \displaystyle -\left(1-\frac{2Gm}{r}{\rm Erf}\left(\frac{M_sr}{2}\right)\right)dt^2+\left(1+\frac{2Gm}{r}{\rm Erf}\left(\frac{M_sr}{2}\right)\right)(dr^2+r^2d\Omega^2)\\
& \displaystyle -4GJ\left[\frac{1}{r}{\rm Erf}\left(\frac{M_sr}{2}\right)-\frac{M_s}{\sqrt{\pi}}e^{-\frac{M_s^2r^2}{4}}\right]{\rm sin}^2\theta d\varphi dt.
\label{metric-IDG}
\end{array}
\end{equation}
From Fig. 1 and 2, it is clear that the metric constructed by using the multipole expansion is a very good approximation to describe the spacetime outside the source, $r>a;$ while in the regime $M_sr\gg2$, we recover the GR predictions, indeed Eq.\eqref{metric-IDG} reduces to the Lense-Thirring metric~\cite{Lense} in Eq.\eqref{lense-thirr}. 
Thus, in the case of ghost-free IDG, for a rotating source we have found a hierarchy of scales: the radius of the source $a$, the Schwarzschild radius $r_{\rm sch}=2Gm$ and the scale of non-locality $r_{NL}\sim2/M_s,$ which have to satisfy the following set of inequalities to preserve the linearity:
\begin{equation}\label{a-ult-inq}
r_{NL}\sim \frac{2}{M_s}>r_{\rm sch}=\frac{2m}{M_p^2}>a\,.
\end{equation}%
As long as the inequality in Eq.\eqref{a-ult-inq} holds, the spacetime metric is valid all the way from $r=\infty$ up to $r=0,$ and it turns out to be free from any curvature singularity, and also devoid of any horizons. Furthermore, since in our case, the $h_{00}$ component is always bounded below unity, there is no ergo-region, as first pointed out in~\cite{Cornell:2017irh}.


\section{Non-Kerr type metric in the full non-linear theory}

We now wish to move towards the full non-linear regime, and show that the Kerr metric does not solve the full non-linear field equations in Eq.\eqref{EOM}. 
First of all, note that strictly speaking the Schwarzschild metric in GR is not a vacuum solution everywhere, indeed there is a delta-Dirac distribution at the origin, so that the stress-energy tensor is non-vanishing at $r=0.$ Thus, even in absence of the Weyl squared term $W_{\mu\nu\rho\sigma}\mathcal{F}_3(\Box_s)W^{\mu\nu\rho\sigma},$ the full non-linear IDG field equations will not allow the Schwarzschild metric as a solution, due to the presence of infinite order covariant derivatives acting on a delta-Dirac source.
We can argue the same also in the case of the Kerr metric. As it was rigorously shown in Ref. \cite{basalin2} by using the theory of distribution, the Kerr metric is not a vacuum solution everywhere but there is a non-vanishing stress-energy tensor expressed as combinations of delta-Dirac and theta-Heaviside on the ring \cite{basalin2}.
%
%
Thus, the infinite order covariant derivatives acting on the theta-Heaviside and delta-Dirac distributions on a ring, generically will generate an object which will have a non-point support. In this sense, the Kerr metric will not pass as a vacuum solution of the IDG field equations.



We now wish to show that the Kerr metric does not pass as a {\it pure} vacuum solution (i.e. $T_{\mu\nu}=0$ everywhere) if the Weyl squared term with a non-constant form-factor (either local or non-local), $\mathcal{F}_{3}(\Box_s)\neq {\it const.},$ is taken into account in the action. Let us check whether the Kerr metric, Eq.\eqref{metric-1}, is a vacuum solution for the full non-linear equations \eqref{EOM}, i.e. whether the Ricci flat condition
\begin{equation}
\mathcal{R}=0,\,\,\,\,\,\,\,\,\,\,\,\,\,\,\,\,\mathcal{R}_{\mu\nu}=0 \label{ricci-flat}
\end{equation} 
satisfy the full non-linear field equations in the vacuum (Eq.(\ref{EOM})), $P_{\alpha\beta}=0,$ whose only non-trivial part that remains to check is
\begin{align}
P^{\alpha\beta}=0=P_{3}^{\alpha\beta}=  &\frac{\alpha_{c}}{8\pi G}\biggl(-g^{\alpha\beta}W^{\mu\nu\lambda\sigma}{\cal F}_{3}(\Box_{s})W_{\mu\nu\lambda\sigma}+4W_{\;\mu\nu\sigma}^{\alpha}{\cal {\cal F}}_{3}(\Box_{s})W^{\beta\mu\nu\sigma}
\nonumber \\
& 
-8\nabla_{\mu}\nabla_{\nu}({\cal {\cal F}}_{3}(\Box_{s})W^{\beta\mu\nu\alpha})-2\Omega_{3}^{\alpha\beta}+g^{\alpha\beta}(\Omega_{3\gamma}^{\;\gamma}+\bar{\Omega}_{3})-8\Delta_{3}^{\alpha\beta}\biggr)\;.\label{P3}
\end{align}
In order to obtain some insight into this problem, let us first consider
the right hand side of $P_{3}^{\alpha\beta}$ up to second order in $\Box_{s}$, namely 
\[
{\cal F}_{3}(\Box_{s})=\left(f_{30}+f_{31}\Box_{s}+f_{32}\Box_{s}^{2}\right),
\]
and study the field equations order by order, as we had done for the static case in Ref.~\cite{Koshelev:2018hpt}.
After some computations (see also Appendix B), we have obtained the following results.
%
	
	\noindent
	{\bf At the zeroth order in $\Box_s$}: This is the case of local fourth order gravity of Stelle~\cite{stelle-1978}:
	\begin{equation}
	S=\frac{1}{ {16\pi G} }\int d^{4}x\sqrt{-g}\left(\mathcal{R}+\alpha_c\left[f_{10}\: \mathcal{R}^2+f_{20}\:\mathcal{R}^{\mu\nu}\mathcal{R}_{\mu\nu}+f_{30}\:W^{\mu\nu\lambda\sigma}W_{\mu\nu\lambda\sigma}\right]\right)\,.\label{stelle-action}
	\end{equation}
	Since we are requiring the condition in Eq.\eqref{ricci-flat}, the full field equations in Eq.\eqref{EOM} are explicitly reduced to Eq.\eqref{P3}, where the only relevant terms that remains to be analyzed is the one corresponding to the form-factor coefficient $f_{30}.$ However, in this case the local contribution from the Weyl squared term with a constant form factor $f_{30}$,  vanishes in 4 dimensions as we can use the Gauss-Bonnet topological invariant to rewrite the Weyl squared in terms of Ricci scalar squared and Ricci tensor squared. Thus, the Kerr metric is still an exact solution for the local fourth order quadratic gravity in Eq.\eqref{stelle-action} \cite{stelle-1978}.
	
	\noindent
	 {\bf At the first order in $\Box_s$}: Even though the Weyl contribution vanishes at zeroth order, this is not the case for the higher powers of box, i.e. $\Box_{s}^{n}$, with $n>0$. Indeed, at the first order in box, we obtain
	\begin{equation}
	P_{3}^{(1)\alpha\beta}(\Box_{s})=\frac{\alpha_{c}}{8\pi G}f_{31}\left(\begin{array}{cccc}
	a_{00} & 0 & 0 & a_{03}\\
	0 & a_{11} & 0 & 0\\
	0 & 0 & a_{22} & 0\\
	a_{30} & 0 & 0 & a_{33}
	\end{array}\right)\:,\label{P-1box}
	\end{equation}
	with the dimensionless matrix elements given by
	\begin{align*}
	a_{00}= & \frac{144G^{2}m^{2}\left(-8a^{4}Gm+a^{2}r\left(100G^{2}m^{2}-8Gmr+5r^{2}\right)+5r^{4}(r-2Gm)\right)}{r^{11}M_{s}^{2}\left(a^{2}+r(r-2Gm)\right)}\;,~
	a_{03}=  -\frac{288aG^{3}m^{3}\left(4a^{2}+25r(r-2Gm)\right)}{r^{11}M_{s}^{2}\left(a^{2}+r(r-2Gm)\right)}\;,
	\end{align*}
	\begin{align*}
	a_{11}= & -\frac{1008G^{2}m^{2}\left(4a^{4}+5a^{2}r(r-2Gm)+r^{2}(r-2Gm)^{2}\right)}{r^{12}M_{s}^{2}}\;,~~~
	a_{22}=  \frac{144G^{2}m^{2}\left(28a^{2}+r(21r-50Gm)\right)}{r^{12}M_{s}^{2}}\;,\\
	a_{30}= & -\frac{288aG^{3}m^{3}\left(4a^{2}+25r(r-2Gm)\right)}{r^{11}M_{s}^{2}\left(a^{2}+r(r-2Gm)\right)}\;,~~~
	a_{33}=  \frac{144G^{2}m^{2}\left(r\left(100G^{2}m^{2}-92Gmr+21r^{2}\right)-8a^{2}Gm\right)}{r^{11}M_{s}^{2}\left(a^{2}+r(r-2Gm)\right)}\;;\label{first-order}
	\end{align*}
	where we have fixed the equatorial plane, $\chi={\rm cos}(\pi/2)=0,$ without any loss of generality. 
	We can also compute the two-rank symmetric tensor $P_3^{\alpha\beta}(\Box_s)$ at higher order in box, see for example Appendix \ref{second-order} for the computations of the second order in box and for the explicit expression of $P_3^{(2)\alpha\beta}(\Box_s)$.

	 {\bf Generic orders in $\Box_s$}: We can now ask what would happen for generic higher orders in $\Box_s$. Note that for the Kerr metric one has
	$\Box_{s}=\frac{1}{M_{s}^{2}\left(a^{2}\chi^{2}+r^{2}\right)}[\left(a^{2}+r(r-2Gm)\right)\partial_{r}^{2}+2(r-Gm)\partial_{r}],$ and by dimensional analysis we can find the behavior of the lowest order in power of $1/r$ at each order in box. We have already seen that the lowest order in $1/r$ at one box goes like $1/r^{10},$ and at two boxes we have $1/r^{12}$; see Appendix \ref{second-order}. By proceeding in the same way, we can notice that at third order in box, the lowest contribution in powers of $1/r$ is
	%
	$f_{33}({G^2m^2}/{r^{14}M_s^6})$,
	%
	and at fourth order in box
	%
	$f_{34}({G^2m^2}/{r^{16}M_s^8})$.
	%
	Finally, we can hint that at $n$-th order in box, the lowest contribution in powers of $1/r$ will be always proportional to
	%
	$f_{3n}({G^2m^2}/{r^{8+2n}M_s^{2n}})$.
	%
By just looking at the lowest order contributions at each order in box, we can notice that the tensor $P_3^{\alpha\beta}$ satisfies the following relation:
\begin{equation}
P_3^{\alpha\beta}\sim f_{31}\mathcal{O}\left(\frac{1}{r^{10}}\right)+f_{32}\mathcal{O}\left(\frac{1}{r^{12}}\right)+\cdots +f_{3n}\mathcal{O}\left(\frac{1}{r^{8+2n}}\right)+\cdots,
\end{equation}
from which it is clear that in order to vanish we would require an unlikely fine-tuning among all coefficients $f_{3n}$. In this respect, Kerr-like metric as in Eq.\eqref{metric-1} cannot be a vacuum solution of the full non-linear field equations in Eq.\eqref{EOM}, indeed it does not pass through at any order in box, $W_{\mu\nu\rho\sigma}\Box^n W^{\mu\nu\rho\sigma}$ with $n\geq 1.$


\section{Conclusions}

Let us briefly conclude our study. In this paper we have studied rotating metric in the case of ghost free IDG~\cite{Biswas:2011ar}. First, we have worked in the linear regime and found the spacetime metric in the case of a stress-energy tensor given by a delta-Dirac distribution on a rotating ring. In GR, this kind of source generates a metric solution which suffers from the presence of a ring singularity, where the Kretschmann scalar blows up, and indeed the metric components diverge on the ring, i.e. for $x^2+y^2=a^2$ and $z=0,$ which mimics the ring singularity appearing in the Kerr metric~\cite{Kerr}. Instead, we have found that in the IDG the spacetime metric turns out to be singularity-free, and  for $r\rightarrow 0$ the metric becomes {\it conformally-flat}, i.e. the cross-term vanishes at the origin, where the metric coincides with the static one~\cite{Biswas:2011ar,Buoninfante:2018rlq}. Moreover, the linear approximation can be trusted all the way from the IR to the UV regime, provided we require slow rotations, $mM_s<M_p^2$, and $a< 2/M_s.$ The last inequality means that the region of non-locality has to engulf the ring-source of radius $a.$ In IDG the angular momentum has to satisfy the inequality $J<(M_p/M_s)^2$ which implies that its value may also exceed one, unlike in GR. We have shown that outside the source, $r>a,$ the spacetime metric can be well described by a multipole expansion which recovers the Lense-Thirring metric in the local regime, $r>2/M_s.$ Finally, we have analyzed the full field equations, and shown that the Kerr metric, seen as Ricci-flat, will not pass as a vacuum solution if the form-factor $\mathcal{F}_3(\Box_s)$ is not constant; indeed, the Weyl contribution does not vanish at each order in box. 

Hence, the notion of rotating blackhole that we have in GR, would be different in IDG, i.e. without singularity, without event horizons, and without ergo region. Indeed, our study might have an interesting impact in astrophysical blackholes, which should be discussed elsewhere in some details. Hopefully, our analysis will also shed some light in presence of LIGO/VIRGO data, and understanding the spacetime near a rotating non-singular compact object.

\section{Acknowledgements}
AM would like to thank Tirthabir Biswas, and Tomi Koivisto for discussions. AM and LB would like to thank Valeri Frolov for very helpful and insightful discussions. LB is thankful to Masahide Yamaguchi for his kind hospitality in Tokyo Institute of Techniology, where part of the work has been carried out. AK and JM are supported by the grant UID/MAT/00212/2013 and COST Action CA15117 (CANTATA). AK is supported by FCT Portugal investigator project IF/01607/2015 and FCT Portugal fellowship SFRH/BPD/105212/2014. AC and GH are supported in part by the National Research Foundation of South Africa.

\appendix

\section{Second order contributions from the Weyl term}\label{second-order}

We now wish to present the explicit expression of the two-rank symmetric tensor $P_3^{\alpha\beta}$ at second order in $\Box_s$. It is given by: 
\begin{equation}
P_{3}^{(2)\alpha\beta}(\Box_{s})=\frac{\alpha_{c}}{8\pi G}f_{32}\left(\begin{array}{cccc}
a_{00} & 0 & 0 & a_{03}\\
0 & a_{11} & 0 & 0\\
0 & 0 & a_{22} & 0\\
a_{30} & 0 & 0 & a_{33}
\end{array}\right)\:.\label{P-2box}
\end{equation}
with the dimensionless matrix elements, defined as 
\begin{align*}
a_{00}= & \dfrac{576G^{2}m^{2}}{r^{15}M_{s}^{4}\left(a^{2}+r(r-2Gm)\right)}\Biggl[4a^{4}Gmr(89Gm-66r)-72a^{6}Gm\\
 & +a^{2}r^{2}\left(-1578G^{3}m^{3}+927G^{2}m^{2}r-656Gmr^{2}+140r^{3}\right)+r^{5}\left(939G^{2}m^{2}-744Gmr+140r^{2}\right)\Biggr]\;,\\
a_{03}= & -\dfrac{1152aG^{3}m^{3}}{r^{15}M_{s}^{4}\left(a^{2}+r(r-2Gm)\right)}\Biggl[36a^{4}-2a^{2}r(89Gm-52r)+r^{2}\left(789G^{2}m^{2}-696Gmr+148r^{2}\right)\Biggr]\;,\\
a_{11}= & \frac{576G^{2}m^{2}}{r^{15}M_{s}^{4}}\Biggl[2a^{4}(193Gm-50r)+a^{2}r\left(-967G^{2}m^{2}+718Gmr-120r^{2}\right)
\end{align*}
\begin{align*}
& +r^{2}\left(390G^{3}m^{3}-459G^{2}m^{2}r+172Gmr^{2}-20r^{3}\right)\Biggr]\;,\\
a_{22}= & \frac{576G^{2}m^{2}\left(a^{2}(100r-426Gm)+r\left(789G^{2}m^{2}-534Gmr+80r^{2}\right)\right)}{r^{15}M_{s}^{4}}\;,\\
a_{30}= & -\frac{1152aG^{3}m^{3}\left(36a^{4}-2a^{2}r(89Gm-52r)+r^{2}\left(789G^{2}m^{2}-696Gmr+148r^{2}\right)\right)}{r^{15}M_{s}^{4}\left(a^{2}+r(r-2Gm)\right)}\;,\\
a_{33}= & -\frac{576G^{2}m^{2}}{r^{15}M_{s}^{4}\left(a^{2}+r(r-2Gm)\right)}\Biggl[72a^{4}Gm-4a^{2}Gmr(89Gm-38r)\\
& +r^{2}\left(1578G^{3}m^{3}-1857G^{2}m^{2}r+694Gmr^{2}-80r^{3}\right)\Biggr]\;.
\end{align*}
%
%



\begin{thebibliography}{1}
	
	\bibitem{-C.-M.}C. M. Will, Living Rev. Rel. 17, 4
	(2014) {[}arXiv:1403.7377 {[}gr-qc{]}{]}.
	
	\bibitem{-B.-P.}B. P. Abbott et al. {[}LIGO Scientific
	and Virgo Collaborations{]}, Phys. Rev. Lett. 116 (2016) no.6, 061102.
	
	\bibitem{Penrose:1964wq} 
  R.~Penrose,
  Phys.\ Rev.\ Lett.\  {\bf 14}, 57 (1965).
  doi:10.1103/PhysRevLett.14.57
  
  \bibitem{Penrose:1969pc} 
  R.~Penrose,
  Riv.\ Nuovo Cim.\  {\bf 1}, 252 (1969)
  [Gen.\ Rel.\ Grav.\  {\bf 34}, 1141 (2002)].
  
\bibitem{Hawking:1969sw} 
  S.~W.~Hawking and R.~Penrose,
  Proc.\ Roy.\ Soc.\ Lond.\ A {\bf 314}, 529 (1970).
	
	\bibitem{Biswas:2005qr} 
	T.~Biswas, A.~Mazumdar and W.~Siegel,
	JCAP {\bf 0603}, 009 (2006)
	[hep-th/0508194].
	
	\bibitem{Biswas:2011ar} 
	T.~Biswas, E.~Gerwick, T.~Koivisto and A.~Mazumdar,
	Phys.\ Rev.\ Lett.\  {\bf 108}, 031101 (2012).
	
	\bibitem{Buoninfante:2018rlq} 
  L.~Buoninfante, A.~S.~Koshelev, G.~Lambiase, J.~Marto and A.~Mazumdar,
  JCAP {\bf 1806}, no. 06, 014 (2018)
  [arXiv:1804.08195 [gr-qc]].
	
	\bibitem{Tseytlin:1995uq} 
	A.~A.~Tseytlin,
	Phys.\ Lett.\ B {\bf 363}, 223 (1995)
	[hep-th/9509050].
	
	\bibitem{Siegel:2003vt}
	W.~Siegel,
	hep-th/0309093.
	
	
	\bibitem{-Yu.-V.}Yu. V. Kuzmin, Yad. Fiz. 50, 1630-1635 (1989).
	
	\bibitem{Tomboulis}
	E. Tomboulis, Phys. Lett. B 97, 77 (1980). E. T. Tomboulis,
	 Renormalization And Asymptotic Freedom In Quantum Gravity, In *Christensen, S.m. ( Ed.): 
	Quantum Theory Of Gravity*, 251-266. E. T. Tomboulis,
	 hep- th/9702146.
	
	\bibitem{Tomboulis:2015gfa} 
  E.~T.~Tomboulis,
  Phys.\ Rev.\ D {\bf 92}, no. 12, 125037 (2015)
  [arXiv:1507.00981 [hep-th]].
P.~Chin and E.~T.~Tomboulis,
  JHEP {\bf 1806} (2018) 014,
  [arXiv:1803.08899 [hep-th]].
	
\bibitem{Talaganis:2014ida} 
  S.~Talaganis, T.~Biswas and A.~Mazumdar,
  Class.\ Quant.\ Grav.\  {\bf 32}, no. 21, 215017 (2015)
  [arXiv:1412.3467 [hep-th]].
	
	
	\bibitem{Witten:1985cc} 
  E.~Witten,
  Nucl.\ Phys.\ B {\bf 268}, 253 (1986).
	
	\bibitem{Witten:2013pra} 
  E.~Witten,
  JHEP {\bf 1504}, 055 (2015)
  [arXiv:1307.5124 [hep-th]].
  
  \bibitem{Woodard}
  D. A. Eliezer and R. P. Woodard, 
  Nucl. Phys. B 325, 389 (1989).
	
	\bibitem{Freund:1987kt} 
  P.~G.~O.~Freund and M.~Olson,
  Phys.\ Lett.\ B {\bf 199}, 186 (1987).
  L.~Brekke, P.~G.~O.~Freund, M.~Olson and E.~Witten,
  Nucl.\ Phys.\ B {\bf 302}, 365 (1988).
  P.~G.~O.~Freund and E.~Witten,
  Phys.\ Lett.\ B {\bf 199}, 191 (1987).
  
  
  \bibitem{Rovelli:2011eq} 
  C.~Rovelli,
  PoS QGQGS {\bf 2011}, 003 (2011)
  [arXiv:1102.3660 [gr-qc]].
  
  
  \bibitem{Ambjorn:2012jv} 
  J.~Ambjorn, A.~Goerlich, J.~Jurkiewicz and R.~Loll,
  Phys.\ Rept.\  {\bf 519}, 127 (2012)
  [arXiv:1203.3591 [hep-th]].
  
  \bibitem{Biswas:2014yia} 
  T.~Biswas and N.~Okada,
  Nucl.\ Phys.\ B {\bf 898}, 113 (2015)
  [arXiv:1407.3331 [hep-ph]].
   A.~Ghoshal, A.~Mazumdar, N.~Okada and D.~Villalba,
  Phys.\ Rev.\ D {\bf 97}, no. 7, 076011 (2018)
  [arXiv:1709.09222 [hep-th]].
  
 \bibitem{Talaganis:2016ovm} 
  S.~Talaganis and A.~Mazumdar,
  Class.\ Quant.\ Grav.\  {\bf 33}, no. 14, 145005 (2016)
  [arXiv:1603.03440 [hep-th]].

\bibitem{Buoninfante:2018mre} 
  L.~Buoninfante, G.~Lambiase and A.~Mazumdar,
  arXiv:1805.03559 [hep-th].


\bibitem{Biswas:2009nx} 
  T.~Biswas, J.~A.~R.~Cembranos and J.~I.~Kapusta,
  Phys.\ Rev.\ Lett.\  {\bf 104}, 021601 (2010)
  [arXiv:0910.2274 [hep-th]].
   T.~Biswas, J.~A.~R.~Cembranos and J.~I.~Kapusta,
  JHEP {\bf 1010}, 048 (2010)
  [arXiv:1005.0430 [hep-th]].
  
  
  \bibitem{Biswas:2013cha} 
	T.~Biswas, A.~Conroy, A.~S.~Koshelev and A.~Mazumdar,
	Class.\ Quant.\ Grav.\  {\bf 31}, 015022 (2014),
	Erratum: [Class.\ Quant.\ Grav.\  {\bf 31}, 159501 (2014)].
	[arXiv:1308.2319 [hep-th]].
	
	
	\bibitem{Buoninfante:2018stt} 
	L.~Buoninfante, G.~Harmsen, S.~Maheshwari and A.~Mazumdar,
	arXiv:1804.09624 [gr-qc].  

\bibitem{Frolov:2015bia} 
	V.~P.~Frolov, A.~Zelnikov and T.~de Paula Netto,
	JHEP {\bf 1506}, 107 (2015)
	[arXiv:1504.00412 [hep-th]].
	
 \bibitem{Frolov}V. P. Frolov,  Phys. Rev. Lett. 115, no. 5, 051102 (2015).
	
	\bibitem{Frolov:2015usa} 
	V.~P.~Frolov and A.~Zelnikov,
	Phys.\ Rev.\ D {\bf 93}, no. 6, 064048 (2016)
	[arXiv:1509.03336 [hep-th]].
	

\bibitem{Koshelev:2018hpt} 
  A.~Koshelev, J.~Marto and A.~Mazumdar,
  arXiv:1803.00309 [gr-qc].

\bibitem{Koshelev:2018rau} 
  A.~S.~Koshelev, J.~Marto and A.~Mazumdar,
  arXiv:1803.07072 [gr-qc].

\bibitem{Biswas:2006bs} 
  T.~Biswas, R.~Brandenberger, A.~Mazumdar and W.~Siegel,
  JCAP {\bf 0712}, 011 (2007)
  [hep-th/0610274].

\bibitem{Edholm:2016hbt} 
	J.~Edholm, A.~S.~Koshelev and A.~Mazumdar,
	Phys.\ Rev.\ D {\bf 94}, no. 10, 104033 (2016).
	
	
	\bibitem{Boos:2018bxf} 
	J.~Boos, V.~P.~Frolov and A.~Zelnikov,
	Phys.\ Rev.\ D {\bf 97}, no. 8, 084021 (2018)
	[arXiv:1802.09573 [gr-qc]].

\bibitem{Kerr}
R. P. Kerr, 
Phys. Rev.Lett. 11, 237 (1963). doi:10.1103/PhysRevLett.11.237


\bibitem{Biswas:2016etb} 
	T.~Biswas, A.~S.~Koshelev and A.~Mazumdar,
	Fundam.\ Theor.\ Phys.\  {\bf 183}, 97 (2016).
	T.~Biswas, A.~S.~Koshelev and A.~Mazumdar,
	Phys.\ Rev.\ D {\bf 95}, no. 4, 043533 (2017).

\bibitem{Woodard-1} 
S. Deser and R. P. Woodard, 
Phys. Rev. Lett. 99, 111301 (2007) [arXiv:0706.2151 [astro-ph]].

\bibitem{Conroy:2014eja} 
  A.~Conroy, T.~Koivisto, A.~Mazumdar and A.~Teimouri,
  Class.\ Quant.\ Grav.\  {\bf 32}, no. 1, 015024 (2015)
  [arXiv:1406.4998 [hep-th]].


  \bibitem{Bravinsky}		
A. O. Barvinsky and G. A. Vilkovisky,  Phys. Rept. 119 (1985) 174;
A. O. Barvinsky, Yu. V. Gusev, V. V. Zhytnikov, and G. A. Vilkovisky,  arXiv:0911.1168 [hep-th];J.~F.~Donoghue and B.~K.~El-Menoufi,
  Phys.\ Rev.\ D {\bf 89}, no. 10, 104062 (2014)
  [arXiv:1402.3252 [gr-qc]].

\bibitem{Buoninfante:2016iuf} 
  L.~Buoninfante,
  arXiv:1610.08744 [gr-qc].
  

  
  
\bibitem{Buoninfante:2018xiw} 
  L.~Buoninfante, A.~S.~Koshelev, G.~Lambiase and A.~Mazumdar,
  arXiv:1802.00399 [gr-qc].
  
  	\bibitem{Giacchini:2018gxp} 
  B.~L.~Giacchini and T.~de Paula Netto,
  arXiv:1806.05664 [gr-qc].

	\bibitem{Koshelev:2017bxd} 
	A.~S.~Koshelev and A.~Mazumdar,
	Phys.\ Rev.\ D {\bf 96}, no. 8, 084069 (2017)
	[arXiv:1707.00273 [gr-qc]].
	
	
	\bibitem{Mathur}
S. D. Mathur, 
Fortsch. Phys. 53 (2005) 793,  [hep-th/0502050]. 
S. D. Mathur, 
Class. Quant. Grav. 26 (2009) 224001, [arXiv:0909.1038 [hep-th]];  
B.~Guo, S.~Hampton and S.~D.~Mathur,
  arXiv:1711.01617 [hep-th].

 \bibitem{Nicolini}
 P. Nicolini, 
 J. Phys. A 38, L631 (2005). P. Nicolini,
A. Smailagic and E. Spallucci, 
 Phys. Lett. B 632, 547
(2006).

  
  \bibitem{Lense}
  H. Thirring, Phys. Zs. 19, 33 [English translation in Gen. Rel. Grav. 16 (1984), 712] (1918). J. Lense, H. Thirring, Phys. Zs. 19:156 (1918). (Translation in Gen. Rel. Grav. 16:727 (1984))
	
	\bibitem{basalin1} H. Basalin, N. Bachbagauer, 
	Phys. Lett. B315 (1993) 93-97, {[}arXiv:gr-qc/9305009{]}.
	
	\bibitem{basalin2} H. Basalin, N. Bachbagauer, 
	Class.Quant.Grav.11:1453-1462,1994, {[}arXiv:gr-qc/9312028{]}.

\bibitem{Cornell:2017irh} 
  A.~S.~Cornell, G.~Harmsen, G.~Lambiase and A.~Mazumdar,
  Phys.\ Rev.\ D {\bf 97}, no. 10, 104006 (2018)
  [arXiv:1710.02162 [gr-qc]].
  
  
\bibitem{Visser:2007fj} 
  M.~Visser,
  arXiv:0706.0622 [gr-qc].
  
  
\bibitem{Hawking:1973uf} 
  S.~W.~Hawking and G.~F.~R.~Ellis,
  "The Large Scale Structure of Space-Time," Cambridge University Press,
  doi:10.1017/CBO9780511524646.
  	
\bibitem{stelle-1978}K. S. Stelle, 
Gen. Rel. Grav. \textbf{9}, 353 (1978).


	
\end{thebibliography}
\end{document}